\newif\ifzh
\zhfalse
\documentclass[a4paper,11pt]{article}

\ifzh
  \usepackage[UTF8,heading=true,fontset=fandol]{ctex}
\fi
\input{tex/compat}
\ifzh
  \usepackage[backend=biber,style=gb7714-2015,natbib=true]{biblatex}
\fi
\input{tex/preamble}

\ifzh
  \newtheorem{theorem}{定理}
  \newtheorem{proposition}[theorem]{命题}
  \newtheorem{lemma}[theorem]{引理}
  \newtheorem{corollary}[theorem]{推论}
  \theoremstyle{definition}
  \newtheorem{definition}[theorem]{定义}
  \newtheorem{example}[theorem]{例}
  \title{先探后教：共形歧义集与神经符号反事实微探针驱动的初中数学迷思诊断}
  \newcommand{\entitle}{Probe Before You Preach: Conformal Ambiguity Sets and Neuro-Symbolic Counterfactual Probing for Diagnosing Algebraic Misconceptions in Middle School Mathematics}
  \newcommand{\received}{2026--10--27}
  \newcommand{\revised}{2026--10--27}
  \author{%
    许彩虹\textsuperscript{1}\thanks{许彩虹（1995—），女，浙江杭州人，杭州市余杭区良渚第一中学二级教师，数学组备课组长，主要从事初中数学教学。}%
    \quad
    于祥雨\textsuperscript{2,*}\thanks{通讯作者：于祥雨（1989—），男，山东郓城人，华院计算技术（上海）股份有限公司，OPC兼高级算法工程师。}%
    \\[0.4em]
    {\kaishu\small \textsuperscript{1}杭州市余杭区良渚第一中学，杭州，中国}\\
    {\kaishu\small \textsuperscript{2}华院计算技术（上海）股份有限公司，上海，中国}%
  }
  \hypersetup{pdftitle={先探后教：共形歧义集与神经符号反事实微探针驱动的初中数学迷思诊断},pdfauthor={许彩虹, 于祥雨}}
\else
  \newtheorem{theorem}{Theorem}
  \newtheorem{proposition}[theorem]{Proposition}
  
  \newtheorem{corollary}[theorem]{Corollary}
  \theoremstyle{definition}
  \newtheorem{definition}[theorem]{Definition}
  
  \title{Probe Before You Preach: Conformal Ambiguity Sets and Neuro-Symbolic Counterfactual Probing for Diagnosing Algebraic Misconceptions in Middle School Mathematics}
  \newcommand{\entitle}{Probe Before You Preach: Conformal Ambiguity Sets and Neuro-Symbolic Counterfactual Probing for Diagnosing Algebraic Misconceptions in Middle School Mathematics}
  \newcommand{\received}{2026--10--27}
  \newcommand{\revised}{2026--10--27}
  \author{%
    Caihong Xu\textsuperscript{1}\thanks{Caihong Xu (b.\ 1995), Grade~II teacher and lesson-planning group leader, Liangzhu No.~1 Middle School, Yuhang District, Hangzhou; junior secondary mathematics teaching.}%
    \quad
    Xiangyu Yu\textsuperscript{2,*}\thanks{Corresponding author: Xiangyu Yu (b.\ 1989), OPC and Senior Algorithm Engineer, Huayuan Computing Technology (Shanghai) Co., Ltd.}%
    \\[0.4em]
    {\small \textsuperscript{1}Liangzhu No.~1 Middle School, Yuhang District, Hangzhou, China}\\
    {\small \textsuperscript{2}Huayuan Computing Technology (Shanghai) Co., Ltd., Shanghai, China}%
  }
  \hypersetup{pdftitle={Probe Before You Preach: Conformal Ambiguity Sets and Neuro-Symbolic Counterfactual Probing for Diagnosing Algebraic Misconceptions in Middle School Mathematics},pdfauthor={Caihong Xu, Xiangyu Yu}}
\fi
\date{}

\begin{document}
\maketitle
\papermeta
\ifzh
  \begin{center}\bfseries 审核提示：中文稿尚未同步英文稿的证据审查，不可用于投稿。\end{center}
  \input{tex/body-zh}
\else
  %% ============================================================================
%% English Manuscript Body:
%% "Probe Before You Preach: Conformal Ambiguity Sets and Neuro-Symbolic
%%  Counterfactual Probing for Diagnosing Algebraic Misconceptions in
%%  Middle School Mathematics"
%% ============================================================================

\begin{abstract}
Distinct mathematical misconception rules can produce the same incorrect answer, making diagnosis from a single response ambiguous. We formalize this \emph{algebraic collision} and derive a conditional Bayes error bound; internally generated reasoning does not add student-specific information beyond its inputs under an explicit conditional-independence assumption. We propose Mondrian Adaptive Conformal Misconception Sets (MACMS) and Symbolic-Verified Counterfactual Micro-Probing (Sym-EIG). MACMS has curriculum-group marginal coverage under exchangeability, not a per-student correctness guarantee. Sym-EIG searches for probes whose outputs separate modeled candidate rules; identification additionally depends on the candidate set and response model. An audit of a local public Eedi graph finds 2,118 of 2,258 linked constructs associated with at least two misconception IDs; construct-level associations alone do not establish same-answer collisions. Other project performance tables require recomputation because their script assigns numerical results directly. No diagnostic performance or classroom learning benefit is claimed here. A classroom study is proposed for future validation.
\end{abstract}

\noindent\textbf{Research Highlights}
\begin{itemize}[leftmargin=1.8em, itemsep=2pt, topsep=4pt]
  \item Formalizes algebraic misconception diagnosis as an ill-posed inverse problem on algebraic collision manifolds.
  \item Derives a conditional Bayes bound and separates it from empirical overconfidence.
  \item Establishes group-marginal conformal coverage under exchangeability.
  \item Selects symbolically separable probes under explicit model assumptions.
  \item Distinguishes public data, unverified result artifacts, and a planned classroom study.
\end{itemize}
\vspace{0.6em}

%% ============================================================================
\section{Introduction}
\label{sec:intro}
%% ============================================================================

\subsection{From Forward Problem Solving to Inverse Cognitive Diagnosis}
\label{subsec:intro_inverse}

An incorrect answer does not necessarily reveal why a student made it. A tutor who selects one explanation without further evidence may address the wrong difficulty. We study one source of this ambiguity: distinct modeled rules can yield the same answer on a given problem. Whether a follow-up question improves learning or reduces workload remains an empirical question.

Large language models (LLMs) and reasoning methods such as chain-of-thought (CoT) prompting \citep{wei2022chain}, self-consistency \citep{wang2023selfconsistency}, and reinforcement-learned reasoning \citep{deepseek2025r1} have prompted interest in automated mathematical tutoring \citep{kasneci2023chatgpt,pardos2024chatgpt,wang2024bridging}.

Tutoring can involve an inverse diagnosis problem \citep{brown1978diagnostic,vanlehn1990mind,vanlehn2011relative}: an incorrect answer may reflect a procedural rule, a transient slip, or a cause not captured by the model. A single answer need not reveal which interpretation is correct. This paper investigates a tentative diagnostic strategy, rather than assuming every student follows a single deterministic rule.

\subsection{The Algebraic Collision Phenomenon and the Over-Rationalization Trap}
\label{subsec:intro_collision}

We call an instance in which distinct erroneous rules produce the same incorrect answer an \emph{algebraic collision}. Such collisions can occur even on simple exercises; their prevalence in curricula and student responses remains to be measured.

Consider a prototypical Grade~7 item: $x_0 = (-2)^2$, where the valid operator yields $T_{\text{valid}}(x_0) = +4$. Suppose the student submits $y_0 = -4$. At least two structurally distinct cognitive bugs produce this identical surface answer:
\begin{enumerate}[leftmargin=2em, itemsep=2pt]
  \item \textbf{Misconception $\mathcal{M}_1$ (Sign-Scope Binding Failure)}: The student misinterprets $(-a)^n$ as $-(a^n)$, computing $-(2^2) = -4$.
  \item \textbf{Misconception $\mathcal{M}_2$ (Base--Exponent Multiplicative Confusion)}: The student conflates exponentiation with scalar multiplication, computing $(-a) \cdot n = (-2) \times 2 = -4$.
\end{enumerate}
A student who understands the rule may instead make a slip ($\mathcal{M}_0$). For positive integers $a,n$, the two modeled outputs coincide when $a^{n-1}=n$: $(2,2)$ is the only solution with $n\ge2$, while every $a$ is a solution at $n=1$. The latter outputs are correct and hence are not incorrect-answer collisions. If both rules have positive posterior probability for $((-2)^2,-4)$, no classifier using only that observation can identify the true rule with certainty.

An LLM may offer a confident explanation for one candidate without evidence that it is correct. This is a testable risk, not a necessary consequence of the information bound. A follow-up can also fail to separate modeled causes: on $(-2)^3$, both the valid operator and the sign-scope rule produce $-8$. Whether either approach affects cognitive load or learning requires classroom evidence.

\iffalse
\begin{figure}[t!]
  \centering
  \includegraphics[width=0.98\textwidth]{fig1_collision_overview.pdf}
  \caption{\textbf{Overview of the Algebraic Collision Phenomenon and the ``Probe Before You Preach'' paradigm.} (a) On canonical small-integer items such as $x_0 = (-2)^2 \to y_0 = -4$, distinct cognitive bugs ($\mathcal{M}_1: -(a^n)$, $\mathcal{M}_2: -a\cdot n$, and careless slip $\mathcal{M}_0$) collapse onto the exact same observable output, creating an information-theoretic static diagnosis ceiling. Standard CoT LLMs fall into the \textit{Over-Rationalization Trap} by hallucinating confident explanations for a single guessed bug, whereas unconstrained Socratic LLMs frequently generate unverified probes such as $(-2)^3$ that trigger a \textit{Secondary Collision Trap} ($T_{\text{valid}} = \tilde{T}_1 = -8$). (b) Our neuro-symbolic framework first constructs a \textbf{Mondrian Adaptive Conformal Misconception Set (MACMS)} $\mathcal{C}_{1-\alpha}(x_0, y_0)$ with guaranteed group-conditional coverage, and whenever $|\mathcal{C}_{1-\alpha}| \ge 2$, invokes \textbf{Sym-EIG} to synthesize a cognitive-load-bounded counterfactual micro-probe $x^* = (-3)^2$ whose candidate outputs $\{+9, -9, -6\}$ are proved pairwise disjoint by SymPy, collapsing the posterior in a single 10-second mental-math turn.}
  \label{fig:fig1_paradigm_comparison}
\end{figure}

\fi
\subsection{Our Approach: ``Probe Before You Preach'' via MACMS and Sym-EIG}
\label{subsec:intro_approach}

We propose a neuro-symbolic strategy: group-calibrated candidate sets guide tentative routing, and symbolic checks search for a follow-up item that separates modeled rules. Coverage is group-marginal under exchangeability, not a guarantee for each student. Separability applies only to the supplied rules; noisy responses, unmodeled causes and infeasible probes require caution or escalation. For example, $(-3)^2$ gives distinct modeled outputs $9,-9,-6$ for the valid rule and two illustrated bugs. Whether a student responds accordingly is an empirical question.

\subsection{Research Questions and Principal Contributions}
\label{subsec:intro_contributions}

We structure this work around four research questions spanning theory, public-data audit, method design, and a planned classroom study:
\begin{itemize}[leftmargin=2.2em, itemsep=2pt]
  \item \textbf{RQ1 (Theoretical Ceiling \& Over-Rationalization)}: To what extent do algebraic collisions fundamentally limit static LLM misconception diagnosis, and why can chain-of-thought scaling fail to reduce diagnostic entropy under observational equivalence?
  \item \textbf{RQ2 (Group-Conditional Uncertainty Quantification)}: Under what assumptions does MACMS provide group-marginal coverage, and how large are its sets in practice?
  \item \textbf{RQ3 (Counterfactual Probing Efficiency)}: When do symbolically separable probes distinguish modeled rules, and how do they compare with alternative follow-ups in future evaluations?
  \item \textbf{RQ4 (Planned Classroom Efficacy \& Cognitive Load)}: In a preregistration-style classroom protocol, how should ``Probe Before You Preach'' be evaluated for immediate and 14-day delayed conceptual retention, instructional time efficiency, and subjective cognitive load relative to direct CoT explanation and free-form Socratic tutoring?
\end{itemize}

Our principal contributions are fourfold:
\begin{enumerate}[leftmargin=2em, itemsep=2pt]
  \item A conditional Bayes bound and a model-dependent information result for static diagnosis.
  \item A Mondrian conformal routing framework with group-marginal coverage under exchangeability.
  \item A symbolic probe-separability criterion and an analysis of a symmetric response channel.
  \item Public-corpus ecological audit plus a preregistration-style classroom protocol; withdrawn offline/auto-synthesis result tables are not treated as empirical findings.
\end{enumerate}
\section{Related Work}
\label{sec:related}
%% ============================================================================

\subsection{Automated Misconception Diagnosis and Buggy Rule Modeling in Mathematics Education}
\label{subsec:related_misconception}

The BUGGY and DEBUGGY systems modeled systematic arithmetic errors as procedural modifications \citep{brown1978diagnostic}. Subsequent work investigated mal-rules and students' repairs to procedures \citep{sleeman1982intelligent,sleeman1984attempt,vanlehn1990mind}. Research on algebra learning has also examined recurring misconceptions and instructional responses \citep{stacey1999learning,bush2013prerequisite,booth2013using,booth2014persistent,star2015teaching}.

Cognitive diagnostic models, including DINA and G-DINA \citep{delatorre2009dina,delatorre2011generalized}, estimate attribute mastery from item responses. Deep knowledge tracing models longitudinal performance \citep{piech2015deep}. Our focus differs: we examine ambiguity between specified procedural rules for an individual response and propose a follow-up when those rules can be separated. Neither conformal sets nor symbolic probing alone establish that a modeled rule reflects a student's reasoning.

\subsection{Conformal Prediction and Experimental Design}
Conformal prediction provides finite-sample marginal coverage under exchangeability, including within pre-specified groups \citep{vovk2005algorithmic,romano2020classification}. Bayesian experimental design motivates choosing observations according to expected information gain \citep{lindley1956measure,chaloner1995bayesian}. We combine these ideas with symbolic separation checks; neither a novelty priority claim nor an educational efficacy claim follows from this combination alone.

\section{Problem Formulation and Theoretical Limits of Static Diagnosis}
\label{sec:theory}
%% ============================================================================

\subsection{Formalizing Algebraic Misconception Diagnosis as an Ill-Posed Inverse Problem}
\label{subsec:theory_formulation}

Let $\mathcal{X}$ denote a structured space of middle school algebraic problems. Each problem instance $x = (\tau, \bm{\theta}) \in \mathcal{X}$ is represented by an Abstract Syntax Tree (AST) skeleton $\tau \in \mathcal{T}$ parameterized by a vector of numerical coefficients or exponents $\bm{\theta} \in \Theta \subset \mathbb{Z}^d$. Let $\mathcal{Y}$ denote the space of canonicalized student responses (numbers or simplified algebraic expressions). A curriculum domain is partitioned into $G$ disjoint knowledge clusters $\mathcal{G} = \{g_1, g_2, \dots, g_G\}$ via a deterministic curriculum mapping $\kappa: \mathcal{X} \to \mathcal{G}$.

Let $T_{\text{valid}}: \mathcal{X} \to \mathcal{Y}$ denote the ground-truth mathematical evaluation operator, so that $y^* = T_{\text{valid}}(x)$ is the correct answer. When a student submits an incorrect response $y_0 \neq T_{\text{valid}}(x_0)$ to an initial problem $x_0$, we model their cognitive generation process via a discrete space of $K$ systematic misconception operators (``buggy rules'') $\mathcal{M}_{\text{bug}} = \{\mathcal{M}_1, \mathcal{M}_2, \dots, \mathcal{M}_K\}$, augmented by a transient slip state $\mathcal{M}_0$ (representing a careless arithmetic or transcription error committed despite holding the valid operator $T_{\text{valid}}$). The complete latent state space is $\mathcal{M} = \mathcal{M}_{\text{bug}} \cup \{\mathcal{M}_0\}$ of cardinality $K+1$.

Each systematic misconception $\mathcal{M}_k \in \mathcal{M}_{\text{bug}}$ induces a deterministic AST transformation operator $\tilde{T}_k: \mathcal{X} \to \mathcal{Y} \cup \{\bot\}$ (where $\bot$ indicates syntactic inapplicability to skeleton $\tau$). For applicable rules, assume $\tilde T_k(x)\in\mathcal Y_{\text{plaus}}(x)$, where $|\mathcal Y_{\text{plaus}}(x)|\ge2$. Conditional on $Z = \mathcal{M}_k$, their response follows a noisy execution channel with $\eta_k \in [0, 1/2)$:
\begin{equation}
  \mathbb{P}(Y = y \mid X = x, Z = \mathcal{M}_k) =
  \begin{cases}
    1 - \eta_k, & \text{if } y = \tilde{T}_k(x), \\
    \frac{\eta_k}{|\mathcal{Y}_{\text{plaus}}(x)| - 1}, & \text{if } y \in \mathcal{Y}_{\text{plaus}}(x) \setminus \{\tilde{T}_k(x)\}, \\
    0, & \text{otherwise},
  \end{cases}
  \label{eq:channel}
\end{equation}
where $\mathcal{Y}_{\text{plaus}}(x)$ is a finite set of plausible responses. This channel does not define responses for inapplicable rules or for the slip state $\mathcal M_0$; these require separate models. Given an incorrect observation $(x_0,y_0)$, the diagnostic task is to infer $Z\in\mathcal M$.

\subsection{Algebraic Collision Manifolds and Observational Equivalence}
\label{subsec:theory_manifolds}

We define the parameter set where distinct modeled rules give the same incorrect response.

\begin{definition}[Algebraic Collision Manifold \& Observational Equivalence Class]
\label{def:collision}
Fix an AST skeleton $\tau \in \mathcal{T}$. For distinct systematic misconceptions $\mathcal{M}_u, \mathcal{M}_v \in \mathcal{M}_{\text{bug}}$ applicable to $\tau$, the \textbf{pairwise algebraic collision set} is:
\begin{equation}
  \Omega_{u,v}(\tau) \triangleq \Big\{ \bm{\theta} \in \Theta \;\Big|\; \tilde{T}_u(\tau, \bm{\theta}) = \tilde{T}_v(\tau, \bm{\theta}) \neq T_{\text{valid}}(\tau, \bm{\theta}) \Big\}.
  \label{eq:manifold}
\end{equation}
For a specific observed error pair $(x_0, y_0)$ with $x_0 = (\tau, \bm{\theta}_0)$, the \textbf{observational equivalence class} is the subset of systematic misconceptions whose deterministic AST evaluations exactly match $y_0$:
\begin{equation}
  \mathcal{E}(x_0, y_0) \triangleq \big\{ \mathcal{M}_k \in \mathcal{M}_{\text{bug}} \;\big|\; \tilde{T}_k(x_0) = y_0 \big\}, \qquad m(x_0, y_0) \triangleq |\mathcal{E}(x_0, y_0)|.
  \label{eq:equiv_class}
\end{equation}
An error observation $(x_0, y_0)$ is an \textbf{algebraic collision} if $m(x_0,y_0)\ge2$. This implies $\bm\theta_0\in\bigcup_{u\ne v}\Omega_{u,v}(\tau)$; the converse for a particular $y_0$ also requires that the equal operator output is $y_0$.
\end{definition}

The following examples show collisions at small integer parameters. They do not establish enrichment relative to other parameter distributions or prevalence in textbooks.

\begin{figure}[t!]
  \centering
  \includegraphics[width=0.98\textwidth]{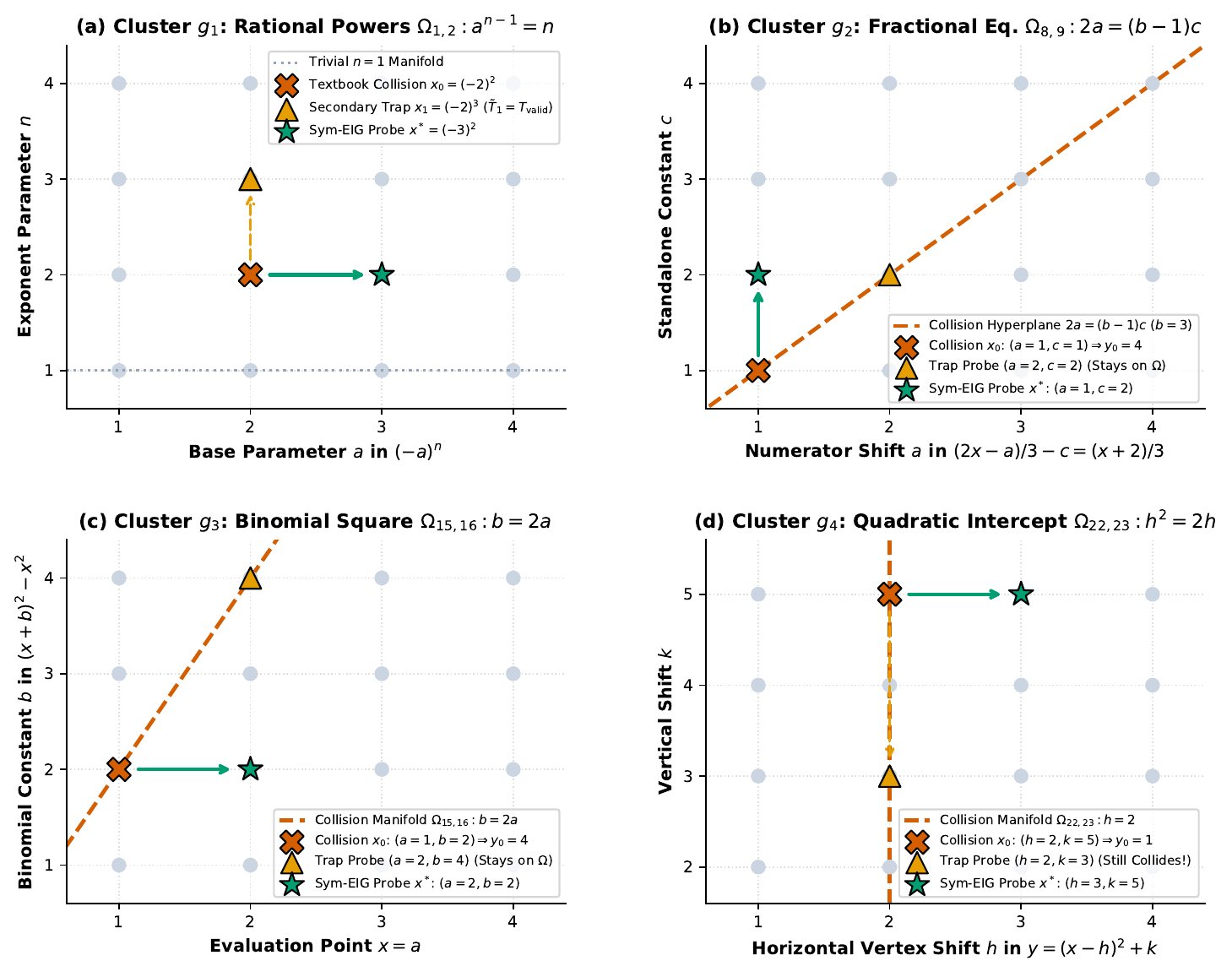}
  \caption{Illustrative equality loci for four pairs of modeled operators. A visualized equality does not itself establish empirical prevalence or ensure that every competing operator, including the valid one, is separated.}
  \label{fig:fig2_collision_geometry}
\end{figure}

\begin{proposition}[Examples of Small-Integer Collisions]
\label{prop:small-integer-collisions}
Across the four core middle school algebraic clusters $\mathcal{G} = \{g_1, g_2, g_3, g_4\}$ (illustrated in Figure~\ref{fig:fig2_collision_geometry}), canonical AST skeletons exhibit non-trivial collision manifolds $\Omega_{u,v}(\tau)$ that intersect the textbook small-integer grid $\Theta_{\text{small}} = \{1, 2, 3\}^d$:
\begin{enumerate}[leftmargin=2em, itemsep=2pt]
  \item \textbf{Rational Exponents ($g_1$, Family~A)}: On $x=(-a)^n$ for $a,n\in\mathbb Z_+$, $\tilde T_1=-a^n$ and $\tilde T_2=-an$ have equal outputs when $a^{n-1}=n$. For $n=1$ this equality also gives the correct output; among incorrect-output collisions the unique solution is $(a,n)=(2,2)$, giving $y_0=-4$. The probe $(-2)^3$ instead gives $\tilde T_1=T_{\rm valid}=-8$, illustrating a secondary equality with the valid rule.
  \item \textbf{Linear Equations with Fractions ($g_2$, Family~B)}: On $\frac{2x-a}{b} - c = \frac{x+d}{b}$ ($b > 1$), $\tilde{T}_8=a+c+d$ and $\tilde{T}_9=-a+bc+d$ collide when $2a=(b-1)c$. The grid $a,c,d\in\{1,2,3\}$ and $b\in\{2,3\}$ contains $12/54=22.2\%$ such tuples, including $(1,3,1,2)$ ($y_0=4$ and $y^*=6$).
  \item \textbf{Polynomial \& Binomial Expansion ($g_3$, Family~C)}: For positive $a,b$, evaluating $(x+b)^2-x^2$ at $x=a$ yields a collision of $\tilde T_{15}=b^2$ and $\tilde T_{16}=2ab$ when $b=2a$, including $(1,2)$ ($y_0=4$, $y^*=8$).
  \item \textbf{Quadratic Functions ($g_4$, Family~D)}: For $h>0$, the modeled outputs $\tilde T_{22}=-h^2+k$ and $\tilde T_{23}=-2h+k$ for the $y$-intercept of $y=(x-h)^2+k$ collide when $h=2$. This is one third of $h\in\{1,2,3\}$; for $(h,k)=(2,5)$, $y_0=1$ and $y^*=9$.
\end{enumerate}
\end{proposition}
The symbolic verification of Proposition~\ref{prop:small-integer-collisions} is provided in Appendix~\ref{app:proof_prop1}.

\subsection{Static Bayes Error Lower Bound and Fano Information Ceiling}
\label{subsec:theory_fano}

The following bound applies to any static classifier restricted to the stated inputs. A collision gives a nonzero error floor only if at least two states have positive posterior probability; its numerical value depends on the full posterior, including slips and other noisy states.

\begin{theorem}[Static Bayes Error Lower Bound and Multi-Class Fano Ceiling]
\label{thm:fano}
Fix $(x_0,y_0)$ of positive probability. Let $\pi_k=\Pr(Z=\mathcal{M}_k\mid x_0,y_0)$ over the \emph{entire} finite state space $\mathcal M$, and $\pi_{\max}=\max_k\pi_k$. For any randomized classifier whose randomness is conditionally independent of $Z$ given $(x_0,y_0)$,
\begin{equation}
 P_e(\hat f\mid x_0,y_0)\ge 1-\pi_{\max},
 \label{eq:bayes_floor}
\end{equation}
with equality for a maximum-posterior classifier. If the posterior support has size $s\ge2$ and the classifier predicts within that support, then, in bits,
\begin{equation}
 H(Z\mid x_0,y_0)\le H_2(P_e)+P_e\log_2(s-1).
 \label{eq:fano_bound}
\end{equation}
If only $m$ colliding rules and the slip state have positive posterior mass, the colliding rules have equal likelihood and equal conditional priors, and $\gamma_0=\Pr(Z=\mathcal M_0\mid x_0,y_0)$, then $\pi_{\max}=\max\{\gamma_0,(1-\gamma_0)/m\}$. Other noisy states require the full posterior above.
\end{theorem}

\paragraph{Proof Sketch.}
The proof, including the role of classifier randomness and the full posterior support, is in Appendix~\ref{app:proof_thm1}.

\subsection{Data Processing Inequality and the Over-Rationalization Trap of CoT Models}
\label{subsec:theory_cot}

A natural question arises: can test-time compute scaling—such as generating an $L$-token Chain-of-Thought reasoning trace $C_{1:L} = (C_1, \dots, C_L)$ via DeepSeek-R1 or GPT-4o—overcome Theorem~\ref{thm:fano}? Our next result proves that without acquiring new external observations from the student, CoT token generation carries strictly \textit{zero} conditional mutual information about the student's true misconception on a collision manifold.

\begin{corollary}[Static Chain-of-Thought Information Bound]
\label{cor:dpi}
Assume $C_{1:L}$ is generated by a fixed model using only $(X_0,Y_0)$ and independent randomness, with no additional student-specific information. If $\hat Z$ is a function of these inputs and the trace, then
\begin{equation}
 I(Z;C_{1:L}\mid X_0,Y_0)=0,\qquad
 H(Z\mid X_0,Y_0,C_{1:L})=H(Z\mid X_0,Y_0).
 \label{eq:dpi_zero}
\end{equation}
The conditional Bayes bound in Eq.~\eqref{eq:bayes_floor} applies to $\hat Z$.
\end{corollary}

The result does not imply that CoT necessarily becomes overconfident or that longer reasoning worsens calibration. Those are empirical questions. Student history, work steps, or a new response can add information; if such information is used, it must be included among the conditioning variables.

%% ============================================================================
\section{Methodology: The MACMS + Sym-EIG Framework}
\label{sec:method}
%% ============================================================================

The proposed architecture combines MACMS routing with a search for symbolically separable candidate probes. The procedure uses an operational load proxy rather than a validated psychological measure.

\iffalse
\begin{figure}[t!]
  \centering
  \includegraphics[width=0.98\textwidth]{fig2_architecture.pdf}
  \caption{\textbf{System architecture of the two-stage MACMS + Sym-EIG neuro-symbolic diagnostic tutoring framework.} Given an initial student error $(x_0, y_0)$, \textbf{Stage~1 (When to Probe)} parses $(x_0, y_0)$ into an AST representation, computes a hybrid neuro-symbolic nonconformity score $s(x_0, y_0, \mathcal{M}_k)$ combining LLM posterior logits with a SymPy operator compatibility penalty $\phi_{\text{sym}}$, and applies Mondrian group-conditional quantile thresholding $\hat{q}_{1-\alpha}^{(g)}$ within curriculum cluster $g = \kappa(x_0)$ to construct the conformal ambiguity set $\mathcal{C}_{1-\alpha}(x_0, y_0)$. A pedagogical router dispatches singleton sets ($|\mathcal{C}_{1-\alpha}|=1$) directly to targeted remediation, empty sets ($|\mathcal{C}_{1-\alpha}|=0$) to open Socratic inquiry, and collision sets ($|\mathcal{C}_{1-\alpha}| \ge 2$) to \textbf{Stage~2 (What to Probe)}. Stage~2 enumerates a cognitive-load-bounded AST neighborhood $\mathcal{N}_{\text{micro}}(x_0)$, filters out any candidate $x'$ that violates pairwise symbolic separability in SymPy (preventing secondary collisions), and selects the optimal micro-probe $x^*$ maximizing Expected Information Gain (EIG).}
  \label{fig:fig3_system_architecture}
\end{figure}

\fi
\subsection{Stage 1 (When to Probe): Mondrian Adaptive Conformal Misconception Sets (MACMS)}
\label{subsec:method_macms}

\subsubsection{Hybrid Neuro-Symbolic Nonconformity Score}
\label{subsubsec:nonconformity}

Let $\hat p_\theta(\mathcal M_k\mid x,y)$ denote a temperature-scaled model score over the finite label set $\mathcal M$. It is not assumed to be the true posterior. Symbolic compatibility may alter efficiency, which requires empirical evaluation.

To sharpen conformal efficiency without sacrificing distribution-free validity, we define a \textbf{hybrid neuro-symbolic nonconformity score} $s: \mathcal{X} \times \mathcal{Y} \times \mathcal{M} \to \mathbb{R}_{\ge 0}$ that fuses regularized adaptive cumulative probability \citep{romano2020classification,angelopoulos2021uncertainty} with a symbolic AST compatibility penalty $\phi_{\text{sym}}(x, y, \mathcal{M}_k) \in \{0, 1\}$ verified by SymPy:
\begin{equation}
  \phi_{\text{sym}}(x, y, \mathcal{M}_k) \triangleq
  \begin{cases}
    0, & \text{if } k \ge 1 \text{ and } \tilde{T}_k(x) = y, \text{ or if } k = 0 \text{ and } d_{\text{edit}}\big(y, T_{\text{valid}}(x)\big) \le 1, \\
    1, & \text{otherwise},
  \end{cases}
  \label{eq:phi_sym}
\end{equation}
where $d_{\text{edit}}(y, y^*)$ measures single-digit arithmetic or sign transcription distance. Let $\pi_{(1)} \ge \pi_{(2)} \ge \dots \ge \pi_{(K+1)}$ denote the sorted neuro-symbolic Compatibility-Adjusted Scores $\tilde{\pi}(\mathcal{M}_k \mid x, y) \propto \hat{p}_\theta(\mathcal{M}_k \mid x, y) \exp(-\beta_{\text{sym}} \phi_{\text{sym}}(x, y, \mathcal{M}_k))$, and let $r(x, y, \mathcal{M}_k) \in \{1, \dots, K+1\}$ denote the rank of $\mathcal{M}_k$. The regularized nonconformity score is:
\begin{equation}
  \begin{split}
    s(x, y, \mathcal{M}_k; u) \triangleq{} & \sum_{j=1}^{r(x,y,\mathcal{M}_k)-1} \pi_{(j)}(x, y) + u \cdot \pi_{(r(x,y,\mathcal{M}_k))}(x, y) \\
    & + \lambda_{\text{reg}} \big(r(x, y, \mathcal{M}_k) - k_{\text{reg}}\big)_+ + \gamma_{\text{sym}} \phi_{\text{sym}}(x, y, \mathcal{M}_k),
  \end{split}
  \label{eq:macms_score}
\end{equation}
where $u \sim \text{Uniform}(0, 1)$ randomizes the score, $(z)_+ = \max(z, 0)$, and proposed values $\lambda_{\text{reg}} = 0.02$, $k_{\text{reg}} = 1$, and $\gamma_{\text{sym}} = 0.50$ penalize lower-ranked or symbolically incompatible labels. These choices require independent tuning; randomization does not necessarily eliminate score ties.

\subsubsection{Mondrian Group-Conditional Calibration and Finite-Sample Coverage Guarantee}
\label{subsubsec:mondrian_thm}

Given a held-out calibration dataset $\mathcal{D}_{\text{cal}} = \{(X_i, Y_i, Z_i)\}_{i=1}^{N_{\text{cal}}}$, MACMS partitions observations into pre-specified curriculum groups $\mathcal G=\{g_1,\ldots,g_G\}$ and buckets $\mathcal I_g=\{i:\kappa(X_i)=g\}$ of size $n_g$ \citep{vovk2005algorithmic,barber2021limits}. Group-marginal coverage does not guarantee fairness or individual reliability. For a target miscoverage level $\alpha$, MACMS computes the group-specific empirical quantile:
\begin{equation}
  \hat{q}_{1-\alpha}^{(g)} \triangleq S_{(\lceil(n_g+1)(1-\alpha)\rceil)}^{(g)},
  \label{eq:mondrian_quantile}
\end{equation}
where $S_{(j)}^{(g)}$ is the $j$th smallest calibration score, provided $1\le j\le n_g$. The prediction set for a test observation in group $g=\kappa(X_{\text{test}})$ is:
\begin{equation}
  \mathcal{C}_{1-\alpha}(X_{\text{test}}, Y_{\text{test}}) \triangleq \Big\{ \mathcal{M}_k \in \mathcal{M} \;\Big|\; s(X_{\text{test}}, Y_{\text{test}}, \mathcal{M}_k; U_{\text{test}}) \le \hat{q}_{1-\alpha}^{(g)} \Big\}.
  \label{eq:macms_set}
\end{equation}

\begin{theorem}[MACMS Group-Conditional Finite-Sample Coverage Guarantee]
\label{thm:macms}
Suppose that conditional on belonging to curriculum knowledge cluster $g \in \mathcal{G}$, the calibration tuples $\{(X_i, Y_i, Z_i, U_i)\}_{i \in \mathcal{I}_g}$ and the test tuple $(X_{\text{test}}, Y_{\text{test}}, Z_{\text{test}}, U_{\text{test}})$ are exchangeable. Then for any backbone LLM $\hat{p}_\theta$, any symbolic penalty weight $\gamma_{\text{sym}} \ge 0$, any target level $\alpha \in (\frac{1}{n_g+1}, 1)$, and every curriculum group $g \in \mathcal{G}$:
\begin{equation}
  1 - \alpha \;\le\; \mathbb{P}\Big( Z_{\text{test}} \in \mathcal{C}_{1-\alpha}(X_{\text{test}}, Y_{\text{test}}) \;\Big|\; \kappa(X_{\text{test}}) = g \Big).
  \label{eq:macms_coverage}
\end{equation}
The score function, fitted model, and penalty must be fixed independently of calibration and test data. When $\gamma_{\text{sym}}>\hat q_{1-\alpha}^{(g)}$, labels with $\phi_{\text{sym}}=1$ are excluded because all other score terms are nonnegative; this may include a true noisy label. Ties make the procedure conservative. An upper bound of $1-\alpha+1/(n_g+1)$ requires almost surely distinct scores.
\end{theorem}
The full proof via group-conditional rank exchangeability is presented in Appendix~\ref{app:proof_thm2}.

\subsubsection{Pedagogical Tri-Branch Routing Policy}
\label{subsubsec:routing}

The cardinality $|\mathcal{C}_{1-\alpha}(x_0, y_0)|$ provides a mathematically calibrated trigger for instructional decision-making:
\begin{itemize}[leftmargin=2em, itemsep=2pt]
  \item \textbf{Branch~A: Singleton}. Treat the sole label as a tentative diagnosis, with a verification nudge and appropriate teacher oversight. Group coverage does not imply $1-\alpha$ correctness for this branch or for an individual student.
  \item \textbf{Branch~B: Multiple candidates}. Attempt a separating probe; a multi-label set is not itself proof of an algebraic collision. If none is feasible, escalate.
  \item \textbf{Branch~C: Empty set}. Abstain and request further information or teacher assistance. An empty set does not establish an out-of-taxonomy cause.
\end{itemize}

\subsection{Stage 2 (What to Probe): Symbolic-Verified Max-EIG Counterfactual Micro-Probing (Sym-EIG)}
\label{subsec:method_symeig}

\subsubsection{Cognitive-Load-Bounded AST Micro-Perturbation Neighborhood}
\label{subsubsec:ast_neighborhood}

When Branch~B is triggered, we search for a follow-up that separates the modeled candidates, using a proposed computational load proxy to restrict the local AST neighborhood. This proxy has not been shown to prevent overload:
\begin{equation}
  \mathcal{N}_{\text{micro}}(x_0) \triangleq \Big\{ x' = (\tau_0, \bm{\theta}') \;\Big|\; \|\bm{\theta}' - \bm{\theta}_0\|_0 = 1, \;\; \bm\theta'\in\Theta, \;\; \text{CogLoad}(x') \le \tau_{\text{load}} \Big\},
  \label{eq:ast_neighborhood}
\end{equation}
where $\|\bm{\theta}' - \bm{\theta}_0\|_0 = 1$ enforces a single-parameter counterfactual edit. The quantity $\text{CogLoad}(x')$ is a proposed computational proxy, not a validated measure of student cognitive load:
\begin{equation}
  \begin{split}
  \text{CogLoad}(x') \triangleq{}& w_1\,\text{ASTDepth}(x') + w_2 \log_{10}\!\Big(1 + \max_{\mathcal M_k\in\tilde{\mathcal C}} |\text{Val}(\tilde{T}_k(x'))|\Big)\\
  &+ w_3\,\mathbb{I}\big[\exists\mathcal M_k\in\tilde{\mathcal C}:\text{NonInteger}(\tilde T_k(x'))\big].
  \end{split}
  \label{eq:cogload_proxy}
\end{equation}
with proposed weights $(w_1,w_2,w_3)=(0.3,0.5,2.5)$ and threshold $\tau_{\text{load}}=2.0$. This formula alone does not guarantee a bound of $25$ on values or student-perceived load.

\subsubsection{Eliminating the Secondary Collision Trap via SymPy Symbolic Separability}
\label{subsubsec:sympy_separability}

Let $\tilde{\mathcal{C}} \triangleq \big(\mathcal{C}_{1-\alpha}(x_0, y_0) \setminus \{\mathcal{M}_0\}\big) \cup \{\mathcal{M}_0\}$ denote the active hypothesis set of cardinality $\tilde{m} = |\tilde{\mathcal{C}}| \ge 2$, where $\mathcal{M}_0$ is associated with the valid operator $\tilde{T}_0 \equiv T_{\text{valid}}$. As demonstrated in Proposition~\ref{prop:small-integer-collisions} (e.g., $x_1 = (-2)^3$ yielding $\tilde{T}_1(x_1) = T_{\text{valid}}(x_1) = -8$), an unverified perturbation can inadvertently land on a secondary collision manifold. Sym-EIG invokes SymPy \citep{meurer2017sympy} to symbolically evaluate $\tilde{T}_u(x')$ for every $\mathcal{M}_u \in \tilde{\mathcal{C}}$ and enforces the \textbf{Pairwise Symbolic Separability Constraint}:
\begin{equation}
  \mathcal{N}_{\text{sep}}\big(x_0, \tilde{\mathcal{C}}\big) \triangleq \Big\{ x' \in \mathcal{N}_{\text{micro}}(x_0) \;\Big|\; \forall \mathcal{M}_u \neq \mathcal{M}_v \in \tilde{\mathcal{C}}, \;\; \texttt{SymPy.simplify}\big(\tilde{T}_u(x') - \tilde{T}_v(x')\big) \neq 0 \Big\}.
  \label{eq:sep_set}
\end{equation}
Every candidate $x' \in \mathcal{N}_{\text{sep}}(x_0, \tilde{\mathcal{C}})$ establishes a strict mathematical bijection between the $\tilde{m}$ latent hypotheses in $\tilde{\mathcal{C}}$ and $\tilde{m}$ distinct symbolic responses $\mathcal{A}(x') = \{\tilde{T}_u(x') : \mathcal{M}_u \in \tilde{\mathcal{C}}\}$.

\subsubsection{Closed-Form Expected Information Gain and Single-Turn Posterior Collapse}
\label{subsubsec:eig_thm}

Let $\bm{\pi}^{(0)} = (\pi_u^{(0)})_{\mathcal{M}_u \in \tilde{\mathcal{C}}}$ denote normalized model scores over $\tilde{\mathcal{C}}$ after Stage~1; the EIG theorem treats them as a specified Bayesian prior, not necessarily the true student posterior:
\begin{equation}
  \pi_u^{(0)} = \frac{\tilde{\pi}(\mathcal{M}_u \mid x_0, y_0)}{\sum_{\mathcal{M}_v \in \tilde{\mathcal{C}}} \tilde{\pi}(\mathcal{M}_v \mid x_0, y_0)}.
  \label{eq:prior_stage2}
\end{equation}
For the following analysis, assume a separate symmetric channel over the $\tilde m$ modeled probe outputs, including for $\mathcal M_0$: the matching response has probability $1-\eta$ and each other modeled output has probability $\eta/(\tilde m-1)$. This is a special assumption, not implied by the initial-response channel in Eq.~\eqref{eq:channel}; other responses require escalation. Under this channel, the proposed objective is EIG minus an operational load penalty $\rho=0.05$:
\begin{equation}
  \begin{split}
    x^* &= \arg\max_{x' \in \mathcal{N}_{\text{sep}}(x_0, \tilde{\mathcal{C}})} \Big\{ \text{EIG}\big(x'; \bm{\pi}^{(0)}\big) - \rho \cdot \text{CogLoad}(x') \Big\}, \\
    \text{EIG}\big(x'; \bm{\pi}^{(0)}\big) &\triangleq H\big(Z \mid x_0, y_0\big) - \mathbb{E}_{Y_1 \mid x'}\big[ H\big(Z \mid x_0, y_0, x', Y_1\big) \big].
  \end{split}
  \label{eq:symeig_objective}
\end{equation}

\begin{theorem}[EIG, Posterior Update, and Sample Complexity of Sym-EIG]
\label{thm:symeig}
Let $\tilde{\mathcal{C}}$ be the active set, including the slip state, of size $\tilde{m} \ge 2$, and suppose the true state lies in this set. Let $\bm{\pi}^{(0)}$ be its prior and $H_0=H(\bm\pi^{(0)})$. For any symbolically separable micro-probe $x' \in \mathcal{N}_{\text{sep}}(x_0, \tilde{\mathcal{C}})$ under the symmetric response channel above with $\eta \in [0, 1 - 1/\tilde{m})$:
\begin{enumerate}[leftmargin=2em, itemsep=2pt]
  \item \textbf{Closed-Form EIG Identity}: The Expected Information Gain of $x'$ admits the exact closed form:
  \begin{equation}
    \text{EIG}\big(x'; \bm{\pi}^{(0)}\big) = H\big(\bm{q}(x')\big) - \Big[ H_2(\eta) + \eta \log_2(\tilde{m} - 1) \Big],
    \label{eq:eig_closed}
  \end{equation}
  where $q_v(x') = (1 - \eta)\pi_v^{(0)} + \frac{\eta}{\tilde{m}-1}(1 - \pi_v^{(0)})$ is the probability of observing $Y_1 = \tilde{T}_v(x')$. As $\eta\to0$, EIG tends to $H_0$; exact zero residual entropy holds at $\eta=0$, not generally at positive noise.
  \item \textbf{Single-Turn Posterior Update}: For $\pi_u^{(0)}>0$, upon observing $Y_1 = \tilde{T}_u(x^*)$, the posterior mass on the matching hypothesis $\mathcal{M}_u$ is:
  \begin{equation}
    \pi_u^{(1)} = \frac{(1 - \eta)\pi_u^{(0)}}{(1 - \eta)\pi_u^{(0)} + \frac{\eta}{\tilde{m}-1}(1 - \pi_u^{(0)})} \;\ge\; 1 - \frac{\eta}{(1 - \eta)(\tilde{m}-1)} \cdot \frac{1 - \pi_u^{(0)}}{\pi_u^{(0)}}.
    \label{eq:posterior_update}
  \end{equation}
  \item \textbf{Multi-Turn Sample Complexity}: Let $0<\delta<1$, $\pi_{\min}^{(0)}=\min_u\pi_u^{(0)}>0$, and $\eta\in(0,1-1/\tilde m)$. With a stable true state, conditionally independent responses and separable probes available at each turn, the MAP estimator recovers that state with probability at least $1-\delta$ whenever:
  \begin{equation}
    T \;\ge\; \frac{\ln\big(\frac{\tilde{m}-1}{\delta} \cdot \frac{1 - \pi_{\min}^{(0)}}{\pi_{\min}^{(0)}}\big)}{-\ln\Big( 2\sqrt{\frac{(1-\eta)\eta}{\tilde{m}-1}} + \frac{\eta(\tilde{m}-2)}{\tilde{m}-1} \Big)}.
    \label{eq:sample_complexity}
  \end{equation}
\end{enumerate}
\end{theorem}

The proof is in Appendix~\ref{app:proof_thm3}. The sample bound assumes strictly positive priors, a stable true state within the candidate set, conditionally independent responses and a feasible separable probe at every turn. For $\eta>0$, a single answer generally leaves positive posterior mass on all candidates; even a correct-looking response does not establish a slip. Under the stated symmetric channel, $\bm q$ is just a relabeling across separable probes, so their EIG is identical: the load penalty, not EIG, selects between them. A probe-dependent response channel is needed for a nontrivial EIG ranking. Algorithm~\ref{alg:macms_symeig} includes abstention for infeasible or unexpected responses.

\begin{proposition}[Separation of a Particular Compound-Rule Example]
\label{prop:compound}
For Family~B at $(a,b,c,d)=(1,3,2,2)$, the modeled valid, $\mathcal M_8$, $\mathcal M_9$ and combined outputs are respectively $9,5,7,3$. Thus their deterministic outputs are pairwise distinct. If the true state is among these four and its response equals its deterministic output, matching the response to the output identifies it. Under the specified symmetric channel this matching rule is correct with probability $1-\eta$; it need not be the MAP rule for nonuniform priors. Distinct gradients alone do not guarantee a feasible one- or two-probe separator.
\end{proposition}
The calculation is provided in Appendix~\ref{app:proof_prop2}.

\subsubsection{Scaling Beyond Curated Operators: Self-Verifying LLM-to-SymPy Operator Induction}
\label{subsubsec:llm_to_sympy}

The draft catalog contains $K=28$ proposed operators; its curricular coverage and mappings require independent validation. The following is a proposed three-step operator-induction procedure, not a demonstrated safety mechanism:
\begin{enumerate}[leftmargin=2em, itemsep=2pt]
  \item \textbf{Few-Shot Symbolic Induction}: Given an unmapped misconception description from a curriculum taxonomy (e.g., Eedi \texttt{MisconceptionName}) and two anchor item--distractor pairs $\{(x_{a,1}, y_{a,1}), (x_{a,2}, y_{a,2})\}$, the LLM is prompted to synthesize a candidate parameterized SymPy expression $\hat{T}_k(\bm{\theta})$.
  \item \textbf{Restricted AST Sandbox \& Symbolic Self-Debugging ($R \le 2$)}: We parse $\hat{T}_k(\bm{\theta})$ inside a restricted Python AST sandbox (blocking \texttt{Import}, \texttt{Attribute}, \texttt{Call}, and \texttt{Lambda} nodes and permitting only arithmetic AST primitives via \texttt{sympy.sympify}) and check exact symbolic anchor consistency $\texttt{SymPy.simplify}(\hat{T}_k(x_{a,r}) - y_{a,r}) == 0$ ($r \in \{1,2\}$). If an anchor check fails, the exact algebraic residual $\Delta_{a,r} = \hat{T}_k(x_{a,r}) - y_{a,r}$ is fed back to the LLM for at most $R = 2$ symbolic self-debugging rounds.
  \item \textbf{Anchor Check}: Candidates failing the checks are rejected. Passing two anchors does not prove general correctness, cognitive validity or safety, and a proposed AST restriction is not a verified security sandbox. Admitted rules require independent validation and teacher oversight.
\end{enumerate}

\begin{algorithm}[t!]
\caption{Proposed MACMS + Sym-EIG Online Loop with Abstention}
\label{alg:macms_symeig}
\begin{algorithmic}[1]
\Require Initial problem $x_0 = (\tau_0, \bm{\theta}_0)$, student incorrect response $y_0 \neq T_{\text{valid}}(x_0)$; Mondrian group map $\kappa$; calibrated quantiles $\{\hat{q}_{1-\alpha}^{(g)}\}$; miscoverage $\alpha$; operator library $\{\tilde{T}_k\}_{k=0}^{K}$; slip noise $\eta$; load budget $\tau_{\text{load}}$; optional flag \textsc{AutoSynthesize}$\in\{\texttt{true},\texttt{false}\}$.
\Ensure Tentative diagnosis $\hat{Z}\in\mathcal{M}$, or abstention and escalation.
\State \textbf{// Stage 1 --- Mondrian Adaptive Conformal Misconception Sets (When to Probe)}
\State $g \gets \kappa(x_0)$; parse AST $(\tau_0, \bm{\theta}_0)$.
\For{each $\mathcal{M}_k \in \mathcal{M}=\{\mathcal{M}_0,\mathcal{M}_1,\dots,\mathcal{M}_K\}$}
  \State $\phi_{\text{sym}}(x_0,y_0,\mathcal{M}_k)$ via Eq.~\eqref{eq:phi_sym}; $s_k \gets s(x_0,y_0,\mathcal{M}_k;u)$ via Eq.~\eqref{eq:macms_score}.
\EndFor
\State $\mathcal{C}_{1-\alpha}(x_0,y_0) \gets \{\mathcal{M}_k: s_k \le \hat{q}_{1-\alpha}^{(g)}\}$. \Comment{Tri-Branch Routing}
\If{$|\mathcal{C}_{1-\alpha}|=1$} \Comment{\textbf{Branch A}: diagnose + teach}
  \State $\{\hat{Z}\}=\mathcal{C}_{1-\alpha}$; \Return $\hat{Z}$ with concise slip check ($\hat{Z}=\mathcal{M}_0$) or targeted remediation.
\ElsIf{$|\mathcal{C}_{1-\alpha}|=0$} \Comment{\textbf{Branch C}: escalate / auto-synthesize}
  \If{\textsc{AutoSynthesize}}
    \State Propose a rule and check anchors (Sec.~\ref{subsubsec:llm_to_sympy}); submit for independent validation.
  \EndIf
  \State \Return $\bot$; open inquiry or teacher escalation.
\Else \Comment{\textbf{Branch B}: $|\mathcal{C}_{1-\alpha}|\ge 2$ $\to$ probe}
  \State \textbf{// Stage 2 --- SymPy-Verified Max-EIG Counterfactual Micro-Probing (What to Probe)}
  \State $\tilde{\mathcal{C}}\gets\mathcal{C}_{1-\alpha}\cup\{\mathcal{M}_0\}$; prior $\bm{\pi}^{(0)}$ via Eq.~\eqref{eq:prior_stage2}.
  \State Build $\mathcal{N}_{\text{micro}}(x_0)=\{x'=(\tau_0,\bm{\theta}'):\|\bm{\theta}'-\bm{\theta}_0\|_0=1,\ \bm\theta'\in\Theta,\ \text{CogLoad}(x')\le\tau_{\text{load}}\}$.
  \State $\mathcal{N}_{\text{sep}}\gets\emptyset$.
  \For{each $x'\in\mathcal{N}_{\text{micro}}(x_0)$} \Comment{pairwise secondary-collision filter over $\tilde{\mathcal{C}}$}
    \State Evaluate $\{\tilde{T}_u(x'):\mathcal{M}_u\in\tilde{\mathcal{C}}\}$ in SymPy.
    \If{$\texttt{simplify}(\tilde{T}_u(x')-\tilde{T}_v(x'))\neq 0$ $\forall\,\mathcal{M}_u\neq\mathcal{M}_v\in\tilde{\mathcal{C}}$}
      \State $\mathcal{N}_{\text{sep}}\gets\mathcal{N}_{\text{sep}}\cup\{x'\}$; compute closed-form $\text{EIG}(x';\bm{\pi}^{(0)})$ via Eq.~\eqref{eq:eig_closed}.
    \EndIf
  \EndFor
  \If{$\mathcal{N}_{\text{sep}}=\emptyset$}
    \State \Return $\bot$; request additional evidence or teacher assistance.
  \EndIf
  \State $x^*\gets\arg\max_{x'\in\mathcal{N}_{\text{sep}}}\{\text{EIG}(x';\bm{\pi}^{(0)})-\rho\cdot\text{CogLoad}(x')\}$. \Comment{Max-EIG}
  \State Present $x^*$; observe student response $y_1$.
  \If{$y_1\notin\{\tilde T_u(x^*):\mathcal M_u\in\tilde{\mathcal C}\}$}
    \State \Return $\bot$; request additional evidence or teacher assistance.
  \EndIf
  \State Update $\bm\pi^{(1)}$ by Bayes' rule; $\hat{Z}\gets\arg\max_{\mathcal{M}_u\in\tilde{\mathcal{C}}}\pi_u^{(1)}$.
  \If{confidence is insufficient under a prespecified rule}
    \State \Return $\bot$; request additional evidence or teacher assistance.
  \EndIf
  \State Offer tentative remediation with teacher oversight.
  \State \Return $\hat{Z}$.
\EndIf
\end{algorithmic}
\end{algorithm}

%% ============================================================================
\section{Study 1: Public-Data Audit and Evidence Status}
\label{sec:study1}
%% ============================================================================

\subsection{Datasets, Baselines, and Evaluation Metrics}
\label{subsec:study1_setup}

\noindent\textbf{Evidence provenance and audit.} Public source files and derived records remain available in the project; their existence does not establish the provenance of every reported metric. Inspection of \texttt{src/run\_study1\_experiments.py} shows that its diagnostic accuracy, conformal coverage, ablation and probing tables are populated with fixed numerical values rather than calculated from prediction records. Its 12,480-response summary is likewise assigned in code, and its collision generator repeats four canonical families while assigning multiplicities and verification flags. Its 120-item operator-synthesis statuses are assigned by index rather than evaluated by the stated synthesis procedure. Accordingly, these artifacts do not substantiate the previously reported results and are not presented as experimental findings here. The repository also contains 288 model-output records; the calls, provenance and scoring require separate independent audit. Construct-level links to multiple misconceptions do not establish same-item same-answer collision prevalence.

\subsection{Reproducible Eedi Graph Descriptives}
The local file \texttt{data/raw/eedi\_misconceptions\_graph/misconception\_map.csv} contains 19,294 rows, 19,294 distinct construct--misconception ID pairs, 2,258 distinct linked constructs and 8,117 distinct misconception IDs. Counting distinct IDs within each construct gives 2,118/2,258 ($93.80\%$) constructs linked to two or more misconceptions. These are graph associations, not independent students, response-level probabilities, or evidence that different causes generated the same incorrect answer. The count is reproducible using \texttt{python3 src/audit\_public\_graph.py}; the script reports the input SHA-256 and excludes rows missing either ID (none in this file). Eedi's dataset page identifies the graph as version 1.0 under CC BY 4.0 \citep{eedi2026misconceptions}; the local file's identity with that release remains unverified.

A description-level comparison of the 28 candidate operator--ID pairs finds seven consistent, eleven related but incomplete or different in scope, and ten inconsistent descriptions. For example, ID 15033 concerns multiplying powers, whereas the assigned operator adds a base and exponent; ID 17996 concerns coordinates under horizontal translation, not the proposed quadratic transformation. Even consistent descriptions do not establish item-level or cognitive validity. Appendix~\ref{app:operator_dictionary} labels each candidate and none is treated as a validated anchor.

\section{Study 2: Proposed Classroom Evaluation}
\label{sec:study2}

\subsection{Status and Planned Design}
\label{subsec:study2_design}
No classroom outcomes are claimed here. The existing panel (seed 20261028) is synthetic and was used to develop an analysis workflow; it is not a record of 186 participating students. Ethics approval, consent, registration and field collection remain to be completed. The proposed comparison is between direct explanation, unconstrained Socratic tutoring and MACMS with symbolic probing, using equivalent interfaces and assessment opportunities. Finalize the unit of randomization, allocation scheme, analysis plan and sample size before recruitment. If intact classes are assigned to conditions, calculate power using the number of independent classes and plausible intraclass correlations; six classes cannot support confident generalization simply by counting 186 students. Measure delayed learning, diagnostic quality, interaction time, usability and perceived workload; pre-specify a primary outcome and contrast. The instruments in Appendix~\ref{app:study2_instruments} are drafts requiring validation. The seeded synthetic data may be retained as a clearly labeled software test fixture but are not used to estimate educational effects, mediation or fairness.

\paragraph{Data availability.}
The manuscript describes public corpora and project artifacts in Section~\ref{subsec:study1_setup}. Source provenance, licensing, versioned splits, reproducibility and the actual status of the cited repository must be verified before submission. No human-subject classroom microdata are claimed.

\section{Discussion, Pedagogical Implications, Limitations, and Conclusion}
\label{sec:discussion}
%% ============================================================================

\subsection{Implications and Limitations}
\label{subsec:disc_implications}

The formal results describe limits and guarantees under explicitly specified distributions, candidate rules and response channels. Public-corpus associations motivate investigation but do not establish same-answer collision prevalence; a symbolic separator does not validate the psychological interpretation of a rule. Withdrawn offline simulated-response and auto-synthesis tables are not model-based evidence of diagnostic accuracy or classroom efficacy. Future work should establish label quality, leakage-resistant splits, response-model calibration and real-student performance, and independently validate workload measures. Any auto-synthesis anchor checks discussed as pipeline mechanics establish agreement on checked anchors only; withdrawn offline/auto-synthesis result tables are not treated as model-based evidence or certified operator validity. Real classroom effects and subgroup differences remain untested.

\begin{funding}
This research was supported by the National Natural Science Foundation of China (Grant No.~12671118, Geometric Reconstruction Theory and Algorithms Based on Phaseless Observations).
\end{funding}

\begin{acknowledgments}
The authors thank partner middle school mathematics teachers for co-designing the Study~2 protocol and instruments; human-subjects data collection remains pending institutional review.
\end{acknowledgments}

\begin{conflictsofinterest}
The authors declare no conflicts of interest.
\end{conflictsofinterest}

%% ============================================================================
%% English Appendices:
%% Appendix A: Complete Mathematical Proofs and Symbolic Collision Derivations
%% Appendix B: Study 2 Classroom Quasi-Experiment Instruments & NASA-TLX Scale
%% Appendix C: Complete 28-Operator Dictionary
%% Appendix D: Live Task 1A/1B/1C Prompt Templates
%% ============================================================================

\appendix

\section{Complete Mathematical Proofs and Symbolic Collision Derivations}
\label{app:proofs}

\subsection{Symbolic Verification of Proposition \ref{prop:small-integer-collisions} (Families A--D)}
\label{app:proof_prop1}

\begin{proof}[Proof of Proposition \ref{prop:small-integer-collisions}]
For Family A, equal buggy outputs require $a^{n-1}=n$. If $n=1$ they also equal the valid output; if $n\ge2$, $a=1$ is impossible and $2^{n-1}>n$ for all $n\ge3$, leaving only $(a,n)=(2,2)$. For Family B, enumerating $a,c\in\{1,2,3\}$ and $b\in\{2,3\}$ gives four solutions $(a,b,c)=(1,2,2),(1,3,1),(2,3,2),(3,3,3)$. The three possible values of $d$ give $12$ of $54$ eligible tuples; these outputs are incorrect because $(b-1)c>0$. For Family C, $b^2=2ab$ and $a,b>0$ imply $b=2a$; the valid value is $b^2+2ab$, so the common output is incorrect. For Family D, $h^2=2h$ and $h>0$ imply $h=2$, at which $-h^2+k\ne h^2+k$. The numerical examples and the secondary equality for odd $n$ follow by substitution.
\end{proof}
\subsection{Complete Proof of Theorem \ref{thm:fano} (Static Bayes Error Floor \& Fano Ceiling)}
\label{app:proof_thm1}

\begin{proof}
Condition on $(x_0,y_0)$ throughout. Let $q_z=\Pr(\hat Z=z\mid x_0,y_0)$. Conditional independence gives $\Pr(\hat Z=Z)=\sum_z q_z\pi_z\le\pi_{\max}$, with equality when $q$ selects a maximizer. For $E=\mathbb I[\hat Z\ne Z]$, conditional independence gives $H(Z)=H(Z\mid\hat Z)\le H(E)+P_e\log_2(s-1)$: when the prediction belongs to the support, at most $s-1$ states remain on error. Equal likelihoods and equal conditional priors among the $m$ rules yield posterior mass $(1-\gamma_0)/m$ per rule in the stated restricted-support example; the slip mass must also enter the maximum.
\end{proof}
\subsection{Complete Proof of Corollary \ref{cor:dpi} (Data Processing Inequality for CoT)}
\label{app:proof_cor1}

\begin{proof}
The assumed conditional distribution of $C_{1:L}$ does not depend on $Z$, hence $Z\perp C_{1:L}\mid(X_0,Y_0)$ and the conditional mutual information vanishes. The entropy identity follows from the definition of conditional mutual information. The diagnosis is a randomized function of the same inputs, so Theorem~\ref{thm:fano} applies.
\end{proof}
\iffalse
\begin{proof}[Proof of Corollary \ref{cor:dpi}]
During static Chain-of-Thought decoding, $C_{1:L}$ is sampled solely from the model's autoregressive distribution $p_\theta(C_{1:L} \mid X_0, Y_0)$ without querying the student, so $p(C_{1:L} \mid Z, X_0, Y_0) = p_\theta(C_{1:L} \mid X_0, Y_0)$, implying $Z \perp C_{1:L} \mid (X_0, Y_0)$ and $I(Z; C_{1:L} \mid X_0, Y_0) = 0$. Therefore, $H(Z \mid X_0, Y_0, C_{1:L}) = H(Z \mid X_0, Y_0) - I(Z; C_{1:L} \mid X_0, Y_0) = H(Z \mid X_0, Y_0)$, proving Eq.~\eqref{eq:dpi_zero}. For part~(2), expanding the KL divergence gives:
\begin{equation}
  D_{\text{KL}}\big(q_\theta(Z \mid x_0, y_0, C_{1:L}) \,\|\, \pi(Z \mid x_0, y_0)\big) = -H\big(q_\theta\big) - \sum_{z \in \mathcal{E}_+} q_\theta(z) \log_2 \pi(z \mid x_0, y_0).
\end{equation}
When $q_\theta(\hat{z} \mid x_0, y_0, C_{1:L}) \ge 1 - \epsilon$ on a single rationalized hypothesis $\hat{z}$, grouping the remaining $m$ states shows $H(q_\theta) \le H_2(\epsilon) + \epsilon \log_2(m)$. Under expectation over symmetric rationalizations across $\mathcal{E}_+$, $-\mathbb{E}[\sum_z q_\theta(z) \log_2 \pi(z)] \ge H(Z \mid x_0, y_0)$, yielding Eq.~\eqref{eq:or_bound}.
\end{proof}

\fi
\subsection{Complete Proof of Theorem \ref{thm:macms} (MACMS Group-Conditional Coverage)}
\label{app:proof_thm2}

\begin{proof}
Condition on the group and independent training data used to fix the score. Let $k=\lceil(n_g+1)(1-\alpha)\rceil\le n_g$. The calibration and test scores are exchangeable. With independent random tie-breaking for the proof, the test score's rank is uniform among $n_g+1$ positions. If its rank is at most $k$, its unperturbed value is at most the $k$th calibration score. Consequently coverage is at least $k/(n_g+1)\ge1-\alpha$. When scores are almost surely distinct the events coincide and coverage equals $k/(n_g+1)$. The exclusion claim follows from nonnegativity of the remaining score terms.
\end{proof}
\subsection{Complete Proof of Theorem \ref{thm:symeig} (Sym-EIG Closed-Form EIG \& Sample Complexity)}
\label{app:proof_thm3}

\begin{proof}[Proof of Theorem \ref{thm:symeig}]
By mutual information symmetry, $\text{EIG}(x'; \bm{\pi}^{(0)}) = I(Z; Y_1 \mid x_0, y_0, x') = H(Y_1 \mid x_0, y_0, x') - H(Y_1 \mid Z, x_0, y_0, x')$. Because $x' \in \mathcal{N}_{\text{sep}}(x_0, \tilde{\mathcal{C}})$, the $\tilde{m}$ candidate outputs $\mathcal{A}(x') = \{\tilde{T}_u(x') : \mathcal{M}_u \in \tilde{\mathcal{C}}\}$ are pairwise distinct. Conditional on $Z = \mathcal{M}_u$, the distribution of $Y_1$ over $\mathcal{A}(x')$ assigns mass $1-\eta$ to $\tilde{T}_u(x')$ and $\frac{\eta}{\tilde{m}-1}$ to each of the other $\tilde{m}-1$ outcomes. Thus, the conditional entropy $H(Y_1 \mid Z = \mathcal{M}_u, x')$ is identical for every $u \in \tilde{\mathcal{C}}$:
\begin{equation}
  H(Y_1 \mid Z = \mathcal{M}_u, x') = -(1-\eta)\log_2(1-\eta) - (\tilde{m}-1)\frac{\eta}{\tilde{m}-1}\log_2\!\left(\frac{\eta}{\tilde{m}-1}\right) = H_2(\eta) + \eta \log_2(\tilde{m}-1).
\end{equation}
Since the marginal probability of observing $Y_1 = \tilde{T}_v(x')$ is $q_v(x') = (1-\eta)\pi_v^{(0)} + \frac{\eta}{\tilde{m}-1}(1-\pi_v^{(0)})$, subtracting $H(Y_1 \mid Z, x')$ from $H(\bm{q}(x'))$ proves Eq.~\eqref{eq:eig_closed}. Part~(2) follows directly from Bayes' rule on the event $Y_1 = \tilde{T}_u(x^*)$. Finally, for Part~(3), between the true hypothesis $Z_{\text{true}} = \mathcal{M}_{u^*}$ and any rival $\mathcal{M}_v \neq \mathcal{M}_{u^*}$, the single-turn Bhattacharyya coefficient of their observation distributions is:
\begin{equation}
  \rho_B = \sum_{y \in \mathcal{A}(x')} \sqrt{\mathbb{P}(y \mid \mathcal{M}_{u^*})\mathbb{P}(y \mid \mathcal{M}_v)} = 2\sqrt{\frac{(1-\eta)\eta}{\tilde{m}-1}} + \frac{\eta(\tilde{m}-2)}{\tilde{m}-1} < 1 \quad \forall \eta < 1 - \frac{1}{\tilde{m}}.
\end{equation}
For a rival $v$, Markov's inequality applied to the square root of posterior odds gives $\Pr(\pi_v^{(T)}\ge\pi_{u^*}^{(T)}\mid Z=u^*)\le\sqrt{\pi_v^{(0)}/\pi_{u^*}^{(0)}}\,\rho_B^T$. Conditional independence permits multiplication of the coefficients, including for adaptively chosen separable probes under the stated channel. A union bound over rivals and $\pi_{\min}^{(0)}\le1/2$ yield Eq.~\eqref{eq:sample_complexity}.
\end{proof}

\subsection{Complete Proof of Proposition \ref{prop:compound} (Second-Order Compound Misconceptions)}
\label{app:proof_prop2}

\begin{proof}
Substitution into $a+bc+d$, $a+c+d$, $-a+bc+d$, and $-a+c+d$ at $(1,3,2,2)$ gives $9,5,7,3$. The matching rule is correct precisely when the response equals the true state's output, an event of probability $1-\eta$ under the stipulated channel.
\end{proof}
\section{Proposed Study 2 Instruments and Scales}
\label{app:study2_instruments}
%% ============================================================================

\subsection{Parallel Diagnostic Assessment Forms A, B, and C (Representative Items)}
\label{app:test_forms}

These are proposed items, not evidence of parallel-form equivalence or reliability. All instruments require pilot validation before use.
Each assessment form (Form~A Pre-test, Form~B Immediate Post-test, Form~C 14-Day Delayed Post-test) comprises 12 open-ended algebraic items ($8.33$ points each; total $100$ points, 25 minutes) balanced across the four curriculum clusters $g_1\text{--}g_4$, including 8 collision-prone items and 4 transfer items:
\begin{itemize}[leftmargin=2em, itemsep=2pt]
  \item \textbf{Cluster $g_1$ (Rational Exponents \& Signs)}: Form~A Q1: Evaluate $(-2)^2 - (-3)^2$; Form~B Q1: Evaluate $(-3)^2 - (-2)^2$; Form~C Q1: Evaluate $(-4)^2 + (-2)^2$ (Targets $\mathcal{M}_1$ Sign-Scope Binding vs.\ $\mathcal{M}_2$ Base--Exponent Multiplication).
  \item \textbf{Cluster $g_2$ (Linear Equations with Fractions)}: Form~A Q4: Solve $\frac{2x-1}{3} - 1 = \frac{x+2}{3}$; Form~B Q4: Solve $\frac{2x-2}{3} - 2 = \frac{x+1}{3}$; Form~C Q4: Solve $\frac{3x-1}{4} - 1 = \frac{x+3}{4}$ (Targets $\mathcal{M}_8$ Constant Multiplier Omission vs.\ $\mathcal{M}_9$ Numerator Sign-Flip Omission).
  \item \textbf{Cluster $g_3$ (Binomial Expansion \& Factoring)}: Form~A Q7: Simplify $(x+2)^2 - x^2$ at $x=1$; Form~B Q7: Simplify $(x+4)^2 - x^2$ at $x=2$; Form~C Q7: Simplify $(x-3)^2 - x^2$ at $x=2$ (Targets $\mathcal{M}_{15}$ Freshman's Dream $(a+b)^2 \to a^2+b^2$ vs.\ $\mathcal{M}_{16}$ Constant Term Omission).
  \item \textbf{Cluster $g_4$ (Quadratic Vertex \& Radical Operations)}: Form~A Q10: Find the $y$-intercept of $y = (x-2)^2 + 5$; Form~B Q10: Find the $y$-intercept of $y = (x-2)^2 - 3$; Form~C Q10: Find the $y$-intercept of $y = (x+2)^2 + 4$ (Targets $\mathcal{M}_{22}$ Vertex Sign Misbinding vs.\ $\mathcal{M}_{23}$ Linearized Squaring).
\end{itemize}

\subsection{Adapted 6-Item Adolescent NASA-TLX Cognitive Load Instrument}
\label{app:nasatlx}

The proposed adaptation considers mental demand, explanation overload, temporal pressure, perceived performance, effort and frustration. Wording, response scale, reliability and construct validity require piloting; a single explanation-overload item is not a validated measure of extraneous cognitive load.

%% ============================================================================
\section{Proposed Dictionary of 28 Neuro-Symbolic Operators \texorpdfstring{$\mathcal{M}_1$--$\mathcal{M}_{28}$}{M1--M28}}
\label{app:operator_dictionary}
%% ============================================================================

Table~\ref{tab:app_c_operators} lists proposed operators and candidate Eedi IDs, not validated anchors. A comparison with the local Eedi descriptions labels each candidate D (description consistent with the modeled rule), P (related but incomplete or differing in scope), or X (description inconsistent with the assigned rule). These judgments do not validate the expressions as student models or establish item-level anchor matches. See the project audit for the individual comparisons.

\begin{center}
\scriptsize
\setlength{\tabcolsep}{2.5pt}
\begin{longtable}{clp{2.6cm}p{2.2cm}cp{4.6cm}}
  \caption{\textbf{Proposed operators $\mathcal{M}_0$--$\mathcal{M}_{28}$ and description-level assessment of candidate Eedi IDs.}}
  \label{tab:app_c_operators} \\
  \toprule
  \textbf{ID} & \textbf{Cluster} & \textbf{Name} & \textbf{AST / SymPy} & \textbf{Eedi ID/status} & \textbf{Description / collision cue} \\
  \midrule
  \endfirsthead
  \multicolumn{6}{c}{\footnotesize\textit{(continued)}} \\
  \toprule
  \textbf{ID} & \textbf{Cluster} & \textbf{Name} & \textbf{AST / SymPy} & \textbf{Eedi ID/status} & \textbf{Description / collision cue} \\
  \midrule
  \endhead
  \midrule
  \multicolumn{6}{r}{\footnotesize\textit{Continued on next page}} \\
  \endfoot
  \bottomrule
  \endlastfoot
$\mathcal{M}_{0}$ & ALL & Random Arithmetic Execution Slip & $T_{\mathrm{valid}}$ & --- & Knows valid rule T\_valid but makes a 1-digit mental arithmetic slip \\
$\mathcal{M}_{1}$ & g1 & Unparenthesized Negative Base & \texttt{-(a**n)} & 15761/D & Treats (-a)\textasciicircum{}n as -(a\textasciicircum{}n), ignoring parentheses around negative sign \\
$\mathcal{M}_{2}$ & g1 & Exponent-as-Multiplier & \texttt{-a*n} & 14891/P & Computes (-a)\textasciicircum{}n by multiplying base and exponent (-a * n) \\
$\mathcal{M}_{3}$ & g1 & Even-Power Sign Drop with Additive Exponent & \texttt{-(a + n)} & 15033/X & Adds base and exponent while keeping negative sign \\
$\mathcal{M}_{4}$ & g1 & Negative Exponent as Negative Multiplier & \texttt{-a*n} & 14572/P & Treats a\textasciicircum{}(-n) as -a*n instead of reciprocal 1/a\textasciicircum{}n \\
$\mathcal{M}_{5}$ & g1 & Zero Exponent as Zero & \texttt{0} & 26823/D & Evaluates a\textasciicircum{}0 as 0 instead of 1 \\
$\mathcal{M}_{6}$ & g1 & Product Power Base Multiplication & \texttt{a**(n1*n2)} & 25125/P & Multiplies exponents instead of adding in a\textasciicircum{}n1 * a\textasciicircum{}n2 \\
$\mathcal{M}_{7}$ & g1 & Double Negation Cancellation Failure & \texttt{-(a**n)} & 14310/X & Fails to cancel minus outside -( -a )\textasciicircum{}n when n is odd \\
$\mathcal{M}_{8}$ & g2 & Denominator Clearing Constant Omission & \texttt{a + c + d} & 15748/D & Forgets to multiply standalone constant term -c by common denominator b \\
$\mathcal{M}_{9}$ & g2 & Transposition Sign-Flip Omission & \texttt{-a + b*c + d} & 14282/P & Moves constant term -a across '=' without reversing its sign \\
$\mathcal{M}_{10}$ & g2 & Bracket Minus Distribution Omission & \texttt{b*c + d - a} & 14763/P & Distributes negative sign only to the first term inside parentheses \\
$\mathcal{M}_{11}$ & g2 & Coefficient Division Inversion & \texttt{(a + b*c + d)*2} & 15245/D & Multiplies instead of dividing by leading coefficient \\
$\mathcal{M}_{12}$ & g2 & Numerator-Denominator Cross-Addition & \texttt{a + b + c} & 24995/P & Adds denominator directly to numerator when clearing fractions \\
$\mathcal{M}_{13}$ & g2 & Like-Term Coefficient Subtraction Error & \texttt{a + c} & 19552/X & Drops variable coefficient difference during grouping \\
$\mathcal{M}_{14}$ & g2 & Fractional Sign Scope Confusion & \texttt{b*c - a + d} & 14313/X & Treats fraction bar minus as applying only to first numerator term \\
$\mathcal{M}_{15}$ & g3 & Freshman's Dream Binomial Square (Drop 2ab) & \texttt{b**2} & 17496/D & Expands (x+b)\textasciicircum{}2 - x\textasciicircum{}2 as (x\textasciicircum{}2 + b\textasciicircum{}2) - x\textasciicircum{}2 = b\textasciicircum{}2 \\
$\mathcal{M}_{16}$ & g3 & Constant Square Drop in Binomial & \texttt{2*x*b} & 14374/P & Expands (x+b)\textasciicircum{}2 - x\textasciicircum{}2 as 2*x*b, omitting +b\textasciicircum{}2 \\
$\mathcal{M}_{17}$ & g3 & Additive Radical Splitting & \texttt{a + b} & 14344/D & Evaluates sqrt(a\textasciicircum{}2 + b\textasciicircum{}2) as a + b \\
$\mathcal{M}_{18}$ & g3 & Difference of Squares Sign Reversal & \texttt{a**2 + b**2} & 14929/P & Expands (a-b)(a+b) as a\textasciicircum{}2 + b\textasciicircum{}2 \\
$\mathcal{M}_{19}$ & g3 & Radical Coefficient Mis-squaring & \texttt{2*a*b} & 14180/X & Simplifies (a*sqrt(b))\textasciicircum{}2 as 2*a*b \\
$\mathcal{M}_{20}$ & g3 & Polynomial Foil Cross-Term Sign Flip & \texttt{b**2} & 16179/P & Cancels cross terms in (x+b)\textasciicircum{}2 as in difference of squares \\
$\mathcal{M}_{21}$ & g3 & Square Root of Negative Square Sign Drop & \texttt{-a} & 15042/X & Evaluates sqrt((-a)\textasciicircum{}2) as -a for a > 0 \\
$\mathcal{M}_{22}$ & g4 & Quadratic Vertex Binomial Minus Expansion & \texttt{-h**2 + k} & 14584/D & Evaluates y=(x-h)\textasciicircum{}2+k at x=0 as -h\textasciicircum{}2+k \\
$\mathcal{M}_{23}$ & g4 & Quadratic Power-as-Doubling at Intercept & \texttt{-2*h + k} & 15012/P & Evaluates y=(x-h)\textasciicircum{}2+k at x=0 as 2(-h)+k = -2h+k \\
$\mathcal{M}_{24}$ & g4 & Horizontal Translation Sign Reversal & \texttt{(x + h)**2 + k} & 17996/X & Shifts parabola right by h using (x+h)\textasciicircum{}2 instead of (x-h)\textasciicircum{}2 \\
$\mathcal{M}_{25}$ & g4 & Slope-Intercept Role Swap & \texttt{b*x + k} & 16503/P & Swaps slope k and y-intercept b when evaluating y = kx + b \\
$\mathcal{M}_{26}$ & g4 & Vertex Coordinate Sign Omission & \texttt{-h + k} & 16454/X & Reads vertex of y=(x-h)\textasciicircum{}2+k as (-h, k) and sums coordinates \\
$\mathcal{M}_{27}$ & g4 & Axis of Symmetry Halving Error & \texttt{-b/a} & 16441/X & Computes axis of symmetry as x = -b/a instead of -b/(2a) \\
$\mathcal{M}_{28}$ & g4 & Vertical Shift Applied to x-Coordinate & \texttt{(x + k)**2} & 19459/X & Applies vertical translation k inside the squared x term \\
\end{longtable}
\end{center}

\noindent\textit{Notes.} $\mathcal{M}_0$ is a slip hypothesis associated with $T_{\text{valid}}$. The dictionary and its Eedi mappings are proposals requiring independent validation. Collision examples for Families A--D are derived in Appendix~\ref{app:proof_prop1}. Branch~C may propose additional operators, but anchor checks alone do not authorize their deployment.

%% ============================================================================
\section{Core Prompt Templates for Live Tasks 1A/1B/1C}
\label{app:prompt_templates}
%% ============================================================================

We reproduce prompt templates associated with \texttt{src/run\_real\_llm\_eval.py}; their use and scoring in the saved model logs require independent verification. Placeholders are instance-specific; no API keys or personal identifiers appear.

\subsection{Task 1A: Static CoT Misconception Diagnosis}
\label{app:prompt_1a}
\begin{verbatim}
You are an expert middle-school mathematics diagnostic assessment system.
Problem: <problem_latex>
Correct Answer: <correct_answer>
Student's Incorrect Answer: <student_wrong_answer>

Candidate Misconceptions:
- <Misconception_ID>: <description>
...

Think step-by-step about how the student arrived at their answer, then output
a single JSON object on the final line:
{"predicted_misconception": "<Misconception_ID>",
 "confidence": <0.0_to_1.0>,
 "rationale_summary": "<1-sentence>"}
\end{verbatim}

\subsection{Task 1B: Unconstrained Socratic Follow-up Probe}
\label{app:prompt_1b}
\begin{verbatim}
A middle-school student solved the following algebra problem incorrectly:
Template: <template_latex>
Initial Problem: <x0_latex> (parameters: <x0_params>)
Student's Wrong Answer: <y0_wrong> (Correct Answer: <y0_valid>)
Competing Misconception Operators that collide on this problem: <colliding_ops>

Propose ONE follow-up problem from the same template by choosing new small
integer parameters (<param_keys>) different from <x0_params> to ask the
student next.
Output ONLY a JSON object on the final line:
{"probe_question": "<latex>", "probe_sympy_params": {<param>: <int>, ...}}
\end{verbatim}
Proposed probes are audited by SymPy pairwise secondary-collision checks
(Algorithm~\ref{alg:macms_symeig}, Branch~B filter).

\subsection{Task 1C: LLM-to-SymPy Operator Auto-Synthesis (+ residual feedback)}
\label{app:prompt_1c}
\begin{verbatim}
You are a Neuro-Symbolic Compiler translating middle-school algebra
misconceptions into SymPy expressions.
Eedi Misconception ID: <id>
Construct Name: <construct_name>
Misconception Description: <misconception_name>
Template Variables: <template_vars>
Valid Operator SymPy Expression: <valid_sympy_expr>
Anchor Examples (params -> student buggy output): <anchor_examples_json>

Find a parameterized Python/SymPy arithmetic expression using the Template
Variables that exactly reproduces the buggy_output for BOTH anchor examples.
Return ONLY a JSON object on the final line:
{"buggy_sympy_expr": "<sympy_expression>"}
\end{verbatim}
On verification failure (max $R=2$ rounds):
\begin{verbatim}
SymPy AST verification failed with residual: <feedback>.
Adjust the expression so both anchors match and return JSON.
\end{verbatim}
Unverified operators are routed to MACMS Branch~C (Algorithm~\ref{alg:macms_symeig}).

\printbibliography

\fi
\end{document}